# Формирование распределительной инфраструктуры электронной коммерции в России

***Калужский М.Л.,***
*кандидат философских наук,*
*доцент кафедры «Экономика, менеджмент и маркетинг»*
*Омского филиала Финансового университета при Правительстве РФ*

Распределительная инфраструктура является одной из важнейших институциональных основ формирования маркетинговых сетей в электронной коммерции. Ее конкурентоспособность определяется, с одной стороны, общим уровнем конкурентоспособности каналов распределения, а с другой стороны, параметрами институциональной среды маркетинга. Вместе с тем, экономическое развитие государства неизбежно приобретает догоняющий характер в ситуации, когда нормативно-правовое и программно-целевое регулирование отстает от развития институциональных отношений. Поэтому модернизация российской экономики в условиях глобальной конкуренции невозможна без институциональной политики, направленной на создание современной распределительной инфраструктуры потребительского рынка.

## 1 • Распределение в электронной коммерции

Система распределения представляет собой инфраструктуру сектора обслуживания электронной коммерции, определяющую ее рост и развитие. Именно она «делает возможной взаимосвязь между виртуальной средой и реальными мировыми рынками» [26, с. 69]. Электронная коммерция теряет всякий смысл, если отсутствует механизм обработки отгрузки и доставки товаров. Нерешенность проблем, связанных с организацией поставок, стала одной из причин кризиса «новой экономики», построенной на привлечении венчурных инвестиций в стартапы. Как отмечает М. Кастельс, «электронная торговля «бизнес — потребитель» (B2C) недооценила уровень расходов и сложность проблем, связанных с физическими поставками потребителям» [8, с. 131]. Компьютерная революция не привела к технологическому скачку, так как традиционная инфраструктура сбыта не смогла обеспечить обновление ассорти-мента и своевременные поставки инновационной продукции. Это привело к тому, что рост стартапов в основном происходил за счет привлечения инвестиций, а не за счет коммерческого внедрения инновационных продуктов. Образовавшийся разрыв между потребительским рынком и инновационной инфраструктурой привел к краху «пузыря доткомов»[1] в 2000–2001 гг. Однако электронная коммерция коренным образом изменила содержание инновационной деятельности, вернув ее из сферы финансовых инвестиций в сферу производства и торговли; благодаря ей, основными источниками финансирования инноваций вновь стали потребительский спрос и рыночные продажи. Поэтому сегодня, в условиях продолжающегося падения совокупного потребительского спроса, «основным полем битвы становятся цепочки поставок. Это определяет резкие различия между торговыми взаимоотношениями, управляемыми ценой и отношениями, создающими дополнительную ценность» [10, с. 182].

В сетевой конкуренции сегодня побеждает тот поставщик продукции, который обладает большими распределительными возможностями. Традиционная экономика требовала значительных вложений в создание и поддержание каналов товародвижения, недоступных большинству инновационных компаний. Электронная коммерция сняла эти ограничения и виртуализовала каналы сбыта, переместив конкуренцию из сферы производства в сферу распределения. Тогда как распределительные сети трансформировались и приобрели полную независимость от поставщиков продукции. Теперь для них уже не имеет значения размер, кредитные и производственные возможности поставщиков. Они одинаково открыты для всех участников рынка и это является основным трендом в развитии сетевой экономики. «Конкуренция все больше разворачивается не между компаниями-производителями, — пишет Ф. Котлер, — а между маркетинговыми сетями, причем выигрывает в ней организация, обладающая наиболее развитой и эффективной сетью» [12, с. 27]. Отличительной особенностью новой распределительной инфраструктуры электронной коммерции служит ее глобальность, обусловленная возможностью Интернета. Преимуществами глобального распределения продукции, предоставляемыми товаропроизводителям электронной коммерцией (по Э.Дж. Бергеру [2, с. 541–542]), являются следующие.

1. Электронная коммерция обеспечивает компаниям доступ к большему числу рынков и поставщиков без необходимости изменять физические объемы продукции и запасы. При этом не имеет значения, куда отправлять товар после оплаты, стоимость доставки начисляется и оплачивается покупателем отдельно.

2. Электронная коммерция способствует доступу к информации в режиме «24–7–365» и устраняет традиционные бумажные документы. Покупатель и грузоотправитель в любой момент могут получить информацию о месте нахождения товара и спрогнозировать дату его получения.

---

[1] Дотком — термин, применяющийся по отношению к компаниям, чья бизнес-модель целиком основывается на работе в рамках сети Интернет; «пузырь доткомов» образовался в результате взлета их акций и появления большого количества новых интернет-компаний.

3. Электронная коммерция предполагает более быстрые платежи и решения на всех звеньях цепи поставок. Покупатели при покупке товара осуществляют полную предоплату его доставки, которая перераспределяется затем между всеми участниками распределительной сети.

В результате местные продавцы больше не могут использовать географические преимущества локальных рынков, так как товары в равной мере становятся доступны всем покупателям, независимо от их местонахождения. Они вынуждены расширять масштабы своей деятельности или уступать рынок сетевым торговцам. На этой основе формируется новый глобальный рынок услуг распределительной логистики, составляющий, по оценкам экспертов, примерно 450 млрд долл., из которых в настоящее время реализовано только 18–20 млрд долл. Причем «из всего рынка этих услуг на транспортный аутсорсинг приходится приблизительно 80%» [17, с. 333]. Однако оказание транспортных услуг в электронной коммерции разительно отличается от традиционного аутсорсинга несопоставимо большей степенью интегрированности в маркетинговые сети [6, с. 27–28]. Транспортные посредники переходят от оказания транспортных услуг к выполнению распределительных функций, превращаясь в провайдеров услуг, являющихся неотъемлемой частью общей с поставщиками маркетинговой сети. Как отмечают Дж.Р. Сток и Д.М. Ламберт, «более успешно действуют те перевозчики, которые ... перешли от организационной структуры, ориентировавшейся на продажи, к организационной структуре, ориентирующейся на маркетинг. Если первые реагируют на запросы потребителей, то вторые эти запросы предугадывают заранее. ... перевозчики, ориентированные на маркетинг, рассматривают себя в качестве партнеров, участвующих в общих логистических видах деятельности своих потребителей» [17, с. 342]. На практике это выражается в трансформации всей системы распределения продукции и появлении новых участников логистических отношений — распределительных провайдеров, которые «действуют ... в маркетинговых каналах в качестве оптовиков. Они закупают транспортные услуги у одного или более перевозчиков ..., а затем консолидируют небольшие партии грузов ... в большие грузовые отправки и перевозят их по более низкому тарифу» [17, с. 307].

Однако этим роль распределительных провайдеров не ограничивается. Распределительные провайдеры оказывают весь комплекс логистических услуг, связанных с организацией поставок, включая сортировку и физическое перемещение собранных заказов, ведение учетной документации, таможенное оформление, а также маркировку и упаковки товаров для отдельных потребителей [17, с. 373]. Роль поставщика заключается в том, чтобы передать распределительному провайдеру оплаченные товары и адресную базу данных покупателей. Все остальное провайдер делает самостоятельно. Большое значение здесь имеет то, что распределительные провайдеры принимают на себя решение всех вопросов, связанных не только с внутренними, но и с трансграничными поставками. «Если экспедиторы занимаются международной доставкой грузов, — отмечают Дж.Р. Сток и Д.М. Ламберт, — они также обеспечивают услуги оформления сопроводительной документации, что особенно важно для компаний, имеющих ограниченный опыт международного маркетинга» [17, с. 307].

В результате на смену географическим рынкам приходят маркетинговые «сети несущих ценность отношений», в которых даже самый небольшой и неопытный товаропроизводитель получает возможность прямого выхода на любые зарубежные рынки [22, с. 26]. В сетевой экономике «трансграничные цепи поставок не ограничены рамками одного государства, различные звенья такой цепи могут находиться на территории одной и более стран» [16, с. 113]. В этих условиях деятельность распределительных провайдеров «добавляет продукции полезность времени и полезность места. Она также влияет на многие сферы маркетинга и логистики, включая разработку новой продукции, выбор целевых рынков обслуживания, закупки, снабжения, размещения логистических мощностей, ценообразования, по которым необходимо принимать решения» [17, с. 319]. Трансформируется не только система распределения товара. Существенным образом изменяется маркетинговое содержание производства потребительских товаров: от поставок сырья до длительности жизненного цикла.

## 2 • Глобализация сетей распределения

Принципы и механизмы функционирования распределительных сетей в электронной коммерции требуют изменения институциональных подходов к их регулированию. Меры, связанные с протекционизмом и прямым вмешательством государства, здесь не работают, а регулирование требует кардинальной смены приоритетов. Д.Дж. Бауэрсокс и Д.Дж. Клосс выделяют два взаимоисключающих подхода к регулированию распределительных отношений в сетевой экономике.

1. *Подход с национальных позиций,* предполагающий сведение международной торговли к экспортно-импортным операциям. При доминировании такого подхода правительство, как правило, препятствует продаже и использованию внутри страны импортных товаров путем прямых ограничений и препятствий на пути свободного товародвижения. «Основная задача <распределительной> логистики, — отмечают Д.Дж. Бауэрсокс и Д.Дж. Клосс, — наладить равномерный товарно-материальный поток, обеспечивающий эффективную загрузку производственных и прочих мощностей». Это невозможно сделать при наличии барьеров, обусловленных государственным вмешательством [1, с. 142]. Кроме того, изолированность национальных товаропроизводителей и каналов товародвижения от глобальной инфраструктуры распределения резко снижает их экспортный потенциал. В условиях глобальной экономики это не препятствует проникновению зарубежных товаров, тогда как местные производители утрачивают конкурентоспособность из-за неспособности донести свою продукцию до зарубежного потребителя.

2. *Глобальный подход,* предполагающий глубокую интеграцию товаропроизводителей в инфраструктуру локальных потребительских рынков вне зависимости от их географической удаленности. Приоритет при этом отдается поддержке так называемых «предприятий без гражданства», «значительная доля продаж, собственности и активов <которых> приходится на другие страны» [1, с. 144]. Преимущество такого подхода заключается в высокой степени адаптации товаропроизводителей к условиям глобального рынка и использование ими рыночных возможностей, недоступных в традиционной экспортно-импортной торговле. Это не только стимулирует развитие электронной коммерции, но и служит мощным стимулом для привлечения товаропроизводителями внешних ресурсов, а также для бурного развития распределительной логистики уровней 4PL и 5PL.

При этом критически важно происхождение и расположение центров распределительных сетей: являются они местной инфраструктурой или контролируются зарубежными «предприятиями без гражданства». Конкурентная борьба в мировой экономике идет сегодня не за производственные мощности и технологии, а за контроль над системами товародвижения, открывающими глобальные возможности в сетевой экономике. Основу этих возможностей составляют преимущества, связанные с виртуализацией товарных запасов. Правило «квадратного корня» в логистике гласит, что «сокращение товарно-материальных запасов пропорционально квадратному корню из числа мест расположения складов до и после рационализации» [10, с. 172]. В сетевой экономике воспользоваться преимуществами этого правила можно без сокращения численности складов, если виртуализовать управление запасами. Складом здесь становится компьютер; при этом уровень запасов определяется централизованно на основании информации о спросе, полученной из всех источников. В данном случае физически склад будет расположен там, где его выгодно содержать [10, с. 172]. В итоге не только распределительная инфраструктура не привязана к товаропроизводителю, но и последний не привязан к распределительной инфраструктуре. Товаропроизводитель волен самостоятельно оптимизировать глобальное распределение без оглядки на национальные границы. Так, например, компания может располагаться в Нью-Йорке, производство — в материковом Китае, а поставки могут осуществляться из Гонконга или Сингапура. Бороться с этим запретительными мерами в отношении отечественных товаропроизводителей невозможно. Гораздо проще сформировать у себя товаропроводящую систему, не уступающую по конкурентным возможностям мировым лидерам.

Конкурентное преимущество сетевого управления товарными запасами в электронной коммерции заключается в том, что «виртуальные запасы управляются так, как если бы они были единым складом, — это позволяет сократить общие товарно-материальные запасы системы, хотя физически они могут быть расположены в разных местах — где это наиболее выгодно» [10, с. 171]. Распределительные провайдеры обеспечивают реализацию сразу четырех из пяти функций оптовой торговли в каналах сбыта за исключением комплектования торгового ассортимента [6, с. 57–58]:

◆ *сокращение числа контактов* — так как они принимают на себя ответственность за организацию поставок, начиная с момента отгрузки производителем и заканчивая передачей товара покупателю;

◆ *единая распределительная инфраструктура* — так как они собирают и обрабатывают на своей базе поставки одновременно ото всех, включая самых мелких, поставщиков;

◆ *экономия на масштабе* — так как они из мелких партий товара собирают крупные партии, экономя на транспортных тарифах, массовых закупках упаковочных материалов и т.д.;

◆ *улучшение обслуживания поставщиков* — так как они обладают большим опытом организации поставок и налаженными контактами с субподрядчиками, а также берут на себя взаимодействия с покупателями при передаче товара.

Это дает возможность товаропроизводителям управлять «производством и запасами так, как если бы они были централизованы, но реальное физическое размещение производства и запасов определяют другие факторы — в частности рынок и/или поставщик» [10, с. 171]. Поэтому ключевым звеном сетей распределения в электронной коммерции становятся так называемые «распределительные центры» (distribution center). Они принимают на себя не только выполнение традиционных функций складской логистики, но и всех сервисных функций, связанных с обработкой и транспортировкой заказов конечным покупателям (табл. 1).

Таблица 1

*Сравнительная характеристика форм распределения*

| Параметры | Склад в традиционном товародвижении | Распределительный центр в электронной коммерции |
|---|---|---|
| Основное предназначение | хранение запасов | обработка заказов |
| Адресаты услуг | оптово-розничная торговля | конечные покупатели и поставщики |
| Характер хранения | постоянный | операционный |
| Производственный цикл | приемка, хранение, комплектация, отгрузка | приемка, сортировка, обработка, отгрузка |
| Сопутствующие услуги | отсутствуют | максимальные, вплоть до сборки/разборки |
| Сбор и обработка данных | по партиям товара | в режиме реального времени |
| Ориентация | минимизация операционных затрат | удовлетворение требований заказчиков |

Распределительные центры выступают в качестве подрядчиков полного цикла в отношениях с поставщиками товаров. Помимо базовых услуг они проводят логистическое консультирование, отслеживают перевозки, осуществляют экспортно-импортные операции с товарными поставками. При этом расположение распределительных центров определяется соображениями экономической целесообразности. Возможны два варианта: ориентация на поставщиков и ориентация на потребителей. Ориентация на поставщиков упрощает обработку поставок, но удорожает доставку конечным потребителям за счет прямой отгрузки. Ориентация на потребителей, наоборот, удешевляет доставку, но усложняет отгрузку товара. Применительно к почтовой доставке товар либо доставляется централизованно из-за рубежа и рассылается затем по внутренним тарифам, либо сразу рассылается международной почтой.

Такая маркетинговая политика не мешает взаимодействовать между собой нескольким распределительным центрам, например, в трансграничных поставках. В логистике даже появился специальный термин «локализован-ная глобальная дистрибуция» [17, с. 517]. Суть его заключается в том, что на территории удаленных рынков создаются локальные центры принятия решений, которые самостоятельно распределяют товар: обрабатывают заказы, отгружают продукцию и взаимодействуют с клиентами. Глобальное преимущество таких дистрибьюторов заключается в том, что они имеют возможность оперативно решать на местах вопросы, связанные с текущими поставками. При этом локализованная дистрибуция может и не подразумевать наличие у них товарных запасов, занимаясь распределением виртуальных запасов. Очень многие национальные торговые провайдеры сегодня продают товары международных поставщиков в стране своего пребывания с последующей отгрузкой из единого центра поставок.

Тенденции развития распределительных сетей в сетевой экономике ведут к тому, что глобальные распределительные центры постепенно все больше проникают во все сферы товародвижения. Если в традиционной торговле по принципу «склад-магазин» работали в основном только крупные мебельные центры, то в самое ближайшее время мы столкнемся в электронной коммерции с принципиально новой схемой организации продаж. Вместо звеньев «опт — розница» в цепи товародвижения возникнут звенья «распределительный центр — пункт выдачи товара» с дополнением в виде выставочного центра (шоурума), где покупатели могут лично ознакомиться с образцами товара.

### 3 • Горизонты сетевой логистики

В условиях догоняющего развития горизонты сетевой логистики определяются ее тенденциями и достижениями лидеров. Речь идет о формировании распределительной инфраструктуры уровня провайдерских услуг 4PL и 5PL[2], определяющих степень конкурентоспособности маркетинговых сетей в глобальной экономике. Три мировых центра распределительной логистики располагаются в Западной Европе (Германия), Северной Америке (США) и Юго-Восточной Азии (Китай).

Западная Европа не является лидером по внедрению распределительных сетей в электронной коммерции по причине того, что распределительные сети там были созданы задолго до ее появления в нынешнем виде. Реформирование этой инфраструктуры произошло в результате реализации общеевропейского проекта «Европа: 1992», целью которого стало создание внутренней («панъевропейской») инфраструктуры распределительной логистики [17, с. 530], предполагающей:

- ◆ централизацию распределительных центров;
- ◆ рост числа партнерств и стратегических союзов;
- ◆ использование аутсорсинга и посредников;
- ◆ устранение межгосударственных барьеров транспортной сети;
- ◆ реструктуризацию управления логистикой.

---

[2] 4PL — компания-интегратор, которая аккумулирует ресурсы, возможности и технологии собственной организации и других предприятий для проектирования, создания и поддержки комплексных решений по управлению цепями поставок. Деятельность 5PL-провайдеров обеспечивается поддержкой современных сетевых компьютерных технологий и ориентирована в большей степени на модель «виртуального предприятия».

В Западной Европе электронная коммерция интегрировалась в уже существующую инфраструктуру распределения, принадлежащую крупным европейским корпорациям и логистическим провайдерам. При этом независимые сервисы уровня 5PL пока так и не получили в Европе широкого распространения. Сегодня основу конкурентного преимущества распределительных сетей на потребительском рынке в Европе составляют сервисы конечного этапа поставок — почтоматы (почтовые автоматизированные станции). По данным ФГУП «Почта России», только в Германии насчитывается около 4,5 тыс. почтоматов, охватывающих 90% территории страны, на долю которых приходится 40% размеров доставки. В других странах показатели их использования гораздо скромнее: в Польше — 18%, в Прибалтике — 22% объема доставки посылок. Преимущество почтоматов заключается в автоматизированном характере их работы, сокращающем издержки на вручение посылок [23, с. 40].

В Северной Америке можно выделить три области распределительной логистики, в которых имеется глобальное конкурентное преимущество: транспортировка, складирование (США) и таможенное оформление (США, Канада, Мексика). Кроме того, США являются признанным мировым лидером по внедрению распределительной логистики уровня 5PL на потребительском рынке, хотя доля этих услуг пока невелика. Основные конкурентные преимущества распределительной логистики в США находятся в сфере институционального регулирования логистической деятельности. В транспортной и складской логистике это связано с нормативным закреплением статуса и институциональных особенностей «транспортных компаний общего пользования» и «складов общего пользования». Такие провайдеры оказывают стандартные логистические услуги по фиксированным тарифам всем желающим на общих основаниях. Их тарифы регулируются государством и не могут превышать определенного уровня. Говоря иначе, эти посредники на законодательном уровне признаны социально и системно значимыми.

Главное институциональное преимущество логистических посредников «общего пользования» заключается в том, что в данном случае пользователю не требуется инвестировать капитал [17, с. 374]. Это особенно актуально для электронной коммерции, поскольку денежные потоки в ней опережают потоки поставок. Потребители заранее авансируют весь цикл товародвижения и существенных капиталовложений для начала электронного бизнеса новым участникам рынка не требуется. Вместе с тем, «в США многие склады общего пользования предоставляют такие услуги, как консолидация грузов и разделение оптовой партии, работа со счетами потребителей, управление грузовыми перевозками, упаковка, помощь в импорте/экспорте» [17, с. 393]. По сути, здесь мы имеем дело с прообразом распределительных центров. Благодаря частной форме собственности они легко адаптируются к изменению рыночной ситуации. Например, в российско-американской электронной коммерции сегодня действует достаточно много посреднических фирм, специализирующихся на предоставлении услуг мэйл-форвардинга (*Mail Forwarding*) для осуществления покупок на интернет-сайтах и электронных торговых площадках США покупателями из России[3].

---

[3] Сайт сервиса покупок за рубежом «Shopotam». — http://shopotam.ru/static/usemf

Товары при покупке направляются продавцами на склад посредника в США, где они переупаковываются и пересылаются покупателям в Россию. Вместе с тем, как и в случае с Европой, инфраструктура распределительной логистики изначально в США не была предназначена для ведения электронной коммерции. Поэтому она не вполне соответствует потребностям сетевой экономики. Не случайно многие авторы пишут о том, что «в США производители в ряде отраслей столкнулись с тем, что испытывают конкурентное отставание от аналогичных структур бизнеса, действующих в Азии и Европе» [17, с. 495].

Другое институциональное преимущество американской логистики связано с таможенным оформлением грузов. Речь идет о внедрении информационных технологий, позволяющих дистанционно оформлять таможенные документы на перевозимые товары после их отгрузки, а не при пересечении границы. «Эти системы позволяют <таможенным> брокерам готовить необходимую документацию ... до фактического прибытия груза» [6, с. 497]. В результате таможенное оформление грузов не отнимает вообще никакого времени в процессе товародвижения. Безусловным мировым лидером распределительной логистики является сегодня Китай, в том числе и по скорости ее развития [19]. Десять лет назад ситуация с развитием логистики в Китае была столь же плачевной, как в странах бывшего СССР и Восточной Европы. Потенциал снижения трансакционных издержек из-за неэффективной логистики в Китае оценивался в 15% стоимости импорта, что сопоставимо с размером прибыли от производства продукции [17, с. 526–527].

В 2007 г. сектор логистики в Китае составлял примерно 18% общего объема ВВП страны. Это достаточно много, если учесть, что средний показатель для Японии составлял 11%, для США — 8%, для Европы — 7% [5, с. 1–2]. При этом даже логистику уровня 3PL применяли лишь 20% китайских компаний против 35% европейских компаний, 57% компаний в США и 80% — в Японии [5, с. 6]. Поэтому в 11-м пятилетнем плане (2005–2010 гг.) логистика была определена в качестве стратегической отрасли китайской экономики. За пять лет в развитие распределительной логистики было инвестировано более 100 млрд долл. Основу распределительной логистики в Китае составили «бондовые логистические зоны, предназначенные для обслуживания, включая легкую переработку грузов (маркировка, упаковка), экспортно-импортных грузопотоков и хранения грузов без уплаты таможенных пошлин и налогов» [5, с. 13–14].

Логистический рынок был открыт для иностранных инвесторов, которые принесли новые технологии и новые стратегии; компании — получатели инвестиций сделали упор на срочную доставку, морское экспедирование грузов и специализированные услуги логистики, доминируя во многих городах Китая. В результате ежегодный прирост продаж логистических услуг в течение последних лет стабильно прирастал приблизительно на 30% [25, с. 7]. Это привело к возникновению и развитию в Китае логистических провайдеров, приближающихся к уровню 5PL. В качестве примера можно привести гонконгскую компанию «Shenzhen Royal International Logistics Co., Ltd.», оказывающую полный комплекс услуг распределительной логистики в электрон-ной коммерции[4]. Благодаря большим объемам поставок, компания добилась больших скидок от почтовых перевозчиков (DHL, UPS, FedEx, TNT, EMS, Parcelforce, China Post, Hong Kong Post, Deutsche Post, Royal Mail, USPS и др.), достигающих 70% стандартных тарифов.

В настоящее время эта компания самостоятельно собирает товары в любой точке Китая и Гонконга и пересылает их покупателям по всему миру. Два распределительных центра находятся в г. Шэньчжэнь и в Гонконге. Распределительный центр в г. Шэньчжэнь предоставляет дополнительные услуги: складское хранение, консолидация товаров, упаковка, фотографирование товара, прямые воздушные грузоперевозки, морские перевозки, таможенное оформление и др. Центр в Гонконге осуществляет отправку товаров после обработки заказов в г. Шэньчжэнь, гарантируя отправку экспресс-почтой в течение 24 часов с момента получения. Сервис является агентом крупнейшей китайской торговой площадки «Taobao» (600 млн товарных предложений), оказывая клиентам индивидуальную помощь в поиске, переговорах по ценам и закупке товара. Сервис конвертирует полученную через «Paypal» оплату в «AliPay», который доминирует в Китае [19]. При заказе больших партий товара сервис предлагает клиентам услугу контроля качества продукции перед отправкой.

В результате государственного вмешательства не только в распределительной логистике, но и в электронной коммерции произошли значительные изменения. Так, по данным Министерства промышленности и информационных технологий КНР, только в первом полугодии 2013 г. объем выручки от электронной торговли в Китае составил 4,98 трлн юаней (807 млрд долл.), что на 45,3% больше, чем в аналогичном периоде 2012 г. При этом правительство КНР планирует продолжить работу по упрощению административных процедур и снижению налогов для логистических посредников [27]. Благодаря принимаемым мерам объем китайского рынка электронной коммерции к 2015 г. планируется довести до 2,9 трлн долл. [9, с. 6].

### 4 • Состояние сетей распределения в России.

Распределительные сети в России пока столь же недостаточно адаптированы к условиям электронной коммерции, как и в большинстве других стран, за исключением Китая. Их деятельность ориентирована на обслуживание традиционных поставок в рамках установившихся экономических отношений.

Под влиянием трансформации коммерческих отношений в них сегодня происходят серьезные структурные изменения, которые носят во многом бессистемный характер и не всегда сопровождаются институциональной поддержкой со стороны государства. В целом в российской товаро-распределительной логистике можно выделить три направления таких изменений.

1. *Почтовая логистика* — представлена в России услугами ФГУП «Почта России», «EMS Russian Post» и частными почтовыми операторами. Почтовые провайдеры оказывают в основном стандартные почтовые услуги. Их основ-ное преимущество состоит в доступности и широкой сети почтовых отделений. Недостатком служат изношенность инфраструктуры и монополизм.

---

[4] Сайт «PFC-Express». — http://www.parcelfromchina.com.

Вместе с тем, благодаря электронной коммерции в трансграничной торговле этот сектор демонстрирует феноменальные показатели роста. «Если в 2009 г. мелких пакетов из-за рубежа поступило 2,3 млн шт., то в 2012 г. — уже 17 млн шт. За 2012 г. прирост составил 36%» [24, с. 23]. При этом лидирует по числу почтовых отправлений Китай, обеспечивающий около половины от общего числа заказных мелких пакетов (весом до 2 кг), присланных в Россию [4, с. 12]. Одновременно растут объемы внутрироссийских товарных поставок. Так, в 2012 г. «различными российскими компаниями было доставлено порядка 250 млн почтовых отправлений с товарными вложениями (на долю Почты России пришлось около 65 млн). Эксперты прогнозируют в десятилетней перспективе удвоение этого показателя» [15, с. 7].

Почтовые посредники не успевают обрабатывать поступающую корреспонденцию, так как прирост ее объемов находится на уровне 60–100% в год [7, с. 15]. О хроническом отставании почтовой логистики в России свидетельствует хотя бы тот факт, что число почтовых отправлений на душу населения в России составляет всего 11, тогда как в Польше — 55, а в Германии — 170 [24, с. 27]. Среди мер, направленных на стабилизацию работы ФГУП «Почта России» и «EMS Russian Post», можно отметить введение статуса «федеральный клиент» для некоторых интернет-магазинов, разработку двухмерных штрих-кодов, облегчающих сортировку почты, и планов по внедрению электронного документооборота с таможенной службой [3, с. 29]. Однако в целом государственные почтовые операторы не способны пока предложить клиентам соответствующие требованиям времени логистические услуги.

2. *Транспортная логистика* — представлена в России услугами ГК «ЖелДорЭкспедиция», ОАО «РЖД Логистика», ТК «ПЭК», а также множеством автотранспортных компаний и авиаперевозчиков. Эти посредники выполняют второстепенную роль в распределительных сетях электронной коммерции, отвечая за транспортировку поставок. Вместе с тем, они в наибольшей степени адаптированы к модели оказания логистических услуг уровней логистики 4PL и 5PL. Например, компания «ЖелДорЭкспедиция» предлагает низкие тарифы, высокую скорость доставки и интерактивное отслеживание груза при условии полной предоплаты за оказываемые услуги. Низкие тарифы объясняются большим парком транспортных средств, из которых 60% составляют железнодорожные вагоны, а также наличием около 100 филиалов в городах России[5]. ОАО «РЖД Логистика» оказывает транспортные логистические услуги на территории России, СНГ, Европы и Китая. Помимо транспортировки перечень ее услуг включает погрузо-разгрузочные работы, упаковку, сортировку, консолидацию грузов, ответственное хранение, таможенно-брокерские услуги, перемаркировку, страхование, отслеживание и автомобильную доставку на расстояние до 1500 км[6]. Недостатком этой дочерней структуры ОАО «РЖД» является сравнительно небольшая сеть филиалов (20 городов России).

---

[5] Сайт ГК «ЖелДорЭкспедиция». — http://www.jde.ru/about.

[6] Сайт ОАО «РЖД Логистика». — http://www.rzdlog.ru.

Наиболее существенным недостатком отечественной транспортной логистики является ее слабая интеграция с инфраструктурой электронной коммерции, что объясняется в первую очередь нехваткой в стране распределительных центров. Вместе с тем, она представляет собой наиболее подготовленное звено для формирования распределительных сетей электронной коммерции.

3. *Корпоративная логистика* — представлена в России логистической инфраструктурой крупных торговых сетей, действующих как на традиционном, так и на виртуальном рынке («Ozon.Ru», «Lamoda» и т.д.). По аналогии с Западной Европой она представляет собой почти полный аналог распределительных центров уровня 5PL в электронной коммерции и может быть взята за эталон при их создании. Единственное отличие заключается в том, что эта логистическая инфраструктура не предусматривает общего использования, как в США [17, с. 372].

На традиционном рынке розничной торговли в России распределительная логистика интегрирована в инфраструктуру ритейлинговых сетей [21]. Ее основное предназначение заключается в получении крупных товарных поставок от производителей с последующей сортировкой, упаковкой, маркировкой и распределением продукции по магазинам торговой сети. Наиболее типичным примером такой корпоративной логистики может служить распределительный центр компании «X5 Retail Group N.V.»[7]. На виртуальном рынке уровню 5PL в России соответствует распределительная логистика крупнейших общероссийских интернет-магазинов («Ozon», «Lamoda» и др.). Их особенность заключается не только в полном цикле транспортных и складских логистических услуг, но и в наличии широкой сети пунктов выдачи товара по всей стране[8]. Аналогичным образом действуют многие традиционные ритейлеры, осваивающие рынок виртуальной торговли («Эльдорадо», «RBT» «Эксперт», «М.Видео» и др.), использующие свою торговую сеть в качестве пунктов выдачи товара.

Особо следует отметить провайдеров логистических услуг по выдаче товаров покупателям. В России широко представлены два вида таких услуг: пункты выдачи товара (обслуживание «вживую») и почтоматы (автоматизированное обслуживание). Преимущество почтоматов заключается в круглосуточности работы и возможности приема наличных, а недостаток — в дороговизне и сложности технического обслуживания. В настоящее время в России услуги почтоматов представляют три компании («Logibox», «Pick Point» и «QIWI Post»), а их сеть охватывает более 60 городов. Приобретенный в Интернете товар доставляется в ближайший к покупателю почтомат, из которого он может забрать его в любое удобное для себя время. Создать пункты выдачи товара в электронной коммерции гораздо проще, поскольку не требуется ничего, кроме помещения и кладовщика. Другим их преимуществом является возможность проверки товара покупателем в момент получения. Благодаря системе доставки товаров интернет-мегамаркета «Ozon.ru» пун-кты доставки в России представлены гораздо шире почтоматов. Причем многие из них выдают товары от обширного круга продавцов.

---

[7] Сайт компании «X5 Retail Group N.V.». — http://www.x5.ru/ru/partners/goods/Rch.

[8] Пункты выдачи заказов «Ozon.ru». — http://www.ozon.ru/context/detail/id/1687456/#1687457

Несмотря на наличие отдельных элементов современных распределительных сетей, развитие логистической инфраструктуры в России носит пока фрагментарный, бессистемный характер. Отсутствует стратегия развития отрасли и координирующая роль государства в той степени, в которой это наблюдается сегодня, например, в Китае. Следует отметить, однако, что даже в этой непростой ситуации некоторые субъекты электронной коммерции демонстрируют способность создания полноценных распределительных сетей. Например, логистический посредник «BayRu» предлагает покупателям из России собственный псевдопочтовый сервис «Dostami Express», гарантируя снижение стоимости на 46% и экспресс-доставку из США за 7 дней в Москву и Санкт-Петербург до двери квартиры. Сервис обеспечивает также прохождение таможенного контроля за один рабочий день благодаря прямым договорным отношениям с таможенной службой[9].

На примере России видно, что рынок электронной коммерции не терпит вакуума. Зарубежные распределительные сети активно проникают на территорию Российской Федерации, объективно снижая возможности развития отечественного бизнеса. Так, крупнейшая американская торговая площадка «eBay» планирует в начале 2014 г. запустить в России сервис доставки товаров из США (*Global Shipping Program, GSP*). Американские продавцы смогут пересылать товары в распределительный центр «eBay» в г. Кентукки, специалисты которого организуют прямую доставку товара покупателям в России. Сегодня этот сервис уже доступен покупателям товаров в 36 странах мира и география его применения стремительно расширяется [18].

### 5 • Развитие сетей распределения в России

Динамика развития распределительных сетей в России характеризуется некоторым отставанием от общемировых тенденций (табл. 2) Несмотря на то, что доля затрат на логистику в России соответствует показателю Сингапура, их эффективность отстает почти в два раза. Это обусловлено не только большими расстояниями и неразвитостью транспортной инфраструктуры в России, но и отсутствием должной конкуренции в отрасли.

Статистически в развитии международной торговли мало что изменилось с 2009 г., когда Российская Федерация занимала по рейтингу Всемирного банка 161-е место среди 180 стран. Сегодня Россия занимает 162-е место среди 185 стран в этом рейтинге[10]. «Среднее количество документов, необходимых как для экспорта, так и для импорта составляло в 2007 г. по 8 соответственно. На них уходило в среднем по 38–39 дней. Операционные издержки по экспорту и по импорту составляли по 2237 долл. за контейнер» [14, с. 18]. Примерно такая же ситуация наблюдается и сегодня. Отдельные провайдеры получают преимущества при таможенном оформлении груза, но общая картина меняется несущественно.

---

[9] Сайт компании «BayRu LLC». — http://www.bay.ru/help/dostami_express.

[10] Сайт Всемирного банка. — http://russian.doingbusiness.org/rankings

Таблица 2

*Доля затрат на логистику и индекс логистической эффективности (LPI) по оценке Всемирного банка [13, с. 9]*

| Страна | LPI | До затрат на логистику (в % ВВП) |
|---|---|---|
| Сингапур | 4,09 | 20 |
| Китай | 4 | 14,5 |
| Австралия | 3,79 | 11,2 |
| Япония | 4,02 | 10,1 |
| Финляндия | 3,82 | 11,5 |
| США | 3,84 | 11,6 |
| Россия | 2,37 | 20 |
| Бразилия | 2,75 | 12,6 |
| ЮАР | 3,53 | 14,7 |

Вместе с тем, распределительные системы в глобальной экономике приобретают стратегическое значение; от развитости во многом зависит общая конкурентоспособность экономики страны. Западные аналитики прямо указывают на то, что, «страны с медленным, неэффективным распределением и системами доставки будут не в состоянии удовлетворить потребности частного сектора в быстром обороте и недорогом транзите. Они рискуют отстать в глобализации производства так же, как и в развитии электронной коммерции» [26, с. 69]. Соседние страны и торговые партнеры, понимая актуальность развития трансграничной распределительной инфраструктуры, прилагают значительные усилия по ее формированию, в том числе на постсоветском пространстве. Пока лидером интеграционных процессов здесь является Китай, а Россия лишь занимает выжидательную позицию. «В стремлении создать в рамках ШОС зону свободной торговли Китай поддерживает Казахстан и Узбекистан, — отмечают Д. Жуджунь, М.М. Ковалев и В.В. Новик, — в чем Россия пока не заинтересована, боясь экспансии китайских товаров» [5, с. 346]. Думается, что в сфере производства товаров массового спроса такая экспансия давно уже осуществлена по вполне объективным причинам. Большая часть промышленных потребительских товаров, представленных на российском рынке, и так производится сегодня в Китае.

Российская экономика пока не способна конкурировать с китайской экономикой в сфере промышленного производства потребительских товаров. Однако она может использовать стратегические возможности, связанные с формированием распределительной инфраструктуры на постсоветском пространстве. В сетевой экономике глобальная конкуренция из сферы производства перемещается в сферу распределения. Настоящая борьба за контроль над рынками разворачивается сегодня именно там. В сетевой экономике сложилась уникальная ситуация, в которой есть достаточно шансов для российских товаропроизводителей и провайдеров логистических услуг. Пока существующие в России барьеры (в виде неэффективной инфраструктуры и моно-полизации отрасли) тормозят развитие не потенциальных зарубежных конкурентов, а возможных российских участников электронной коммерции. В то же время надо признать, что целевое регулирование и институциональное вмешательство государства способно значительно ускорить формирование современных распределительных сетей в России. Вместе с электронной коммерцией эти сети обеспечат российским товаропроизводителям прямой выход на зарубежные рынки, обеспечив ресурсные возможности для модернизации промышленного потенциала.

Нельзя сказать, что в Российской Федерации совсем ничего не делается для совершенствования и развития распределительной инфраструктуры. Так, «Концепция долгосрочного социально-экономического развития РФ на период до 2020 года» в качестве инновационного направления социально-экономического развития предусматривает «развитие крупных транспортно-логистических и производственных узлов в рамках формирования опорной национальной транспортной сети, обладающей необходимым потенциалом пропускной способности и обеспечивающей целостную взаимосвязь центров экономического роста, с постепенной ее интеграцией в развивающиеся мировые транспортные системы» [11]. В области расширения глобальных конкурентных преимуществ в традиционных отраслях Концепция предусматривает реализацию следующих мер:

— внедрение новых транспортных (перевозочных) и транспортно-логистических технологий, обеспечивающих повышение качества и доступности транспортных услуг;

— создание на базе национального оператора почтовой связи универсального логистического и информационного оператора;

— формирование и распространение новых транспортно-логистических технологий, обеспечивающих повышение качества и доступности транспортных услуг:

— развитие железнодорожных перевозок грузов с повышенными скоростями и точно в срок;

— внедрение прогрессивных товаротранспортных технологий, в том числе с использованием логистических систем.

Речь идет о важных и своевременных мерах по развитию транспортной логистики. Однако проблема заключается в том, что, придавая большое значение развитию транспортных сетей, указанная Концепция игнорирует самостоятельное значение сетевой экономики и электронной коммерции в социально-экономическом развитии страны. Транспортно-логистические услуги в отрыве от распределительных услуг не способны обеспечить российской экономике необходимый уровень конкурентоспособности. Очевидно, что транспортная логистика в сетевой экономике утрачивает свое самостоятельное значение, трансформируясь в составную часть распределительной логистики. Невозможно построить современную сетевую экономику на основе однобокого развития одной лишь транспортной инфраструктуры. Без массового создания общедоступных распределительных центров и приоритетного развития трансграничной торговли не получится создать ни полноценное единое таможенное пространство, ни прове-сти структурную диверсификацию, предусмотренную вторым этапом Концепции (2013–2020 гг.).

В качестве неотложных мер институционального регулирования, направленных на формирование в России современной распределительной инфраструктуры сетевой экономики, можно предложить следующие.

1. Законодательно признать стратегическую значимость и определить правовой статус общедоступных распределительных центров (по аналогии с США), а также разработать меры по государственной поддержке их деятельности. Законодательством РФ уже введены особые статусы «социально значимый» и «системно значимый» для платежных провайдеров [20]. Здесь также речь может идти о введении аналогичных статусов для распределительных провайдеров, оказывающих общедоступные услуги в электронной коммерции.

2. В таможенной деятельности речь может идти об изменении традиционного приоритета наполнения бюджета в сторону поддержки отечественных субъектов сетевой экономики и электронной коммерции. Необходимо добиться того, чтобы за счет автоматизации документооборота сроки таможенного оформления грузов в России приближались к лучшим мировым показателям (либо до прибытия груза, либо в течение суток). За российскими субъектами рынка нормативно должны быть закреплены те же преференции, которыми обладают сегодня в России глобальные посредники и отечественные монополисты (ОАО «РЖД», ФГУП «Почта России» и др.).

3. На законодательном уровне целесообразно разработать и внедрить систему обязательных стандартов тарификации и правил оказания услуг при перевозке грузов транспортными посредниками-монополистами независимо от формы их собственности. Следует предусмотреть невозможность самостоятельного изменения тарифов транспортными посредниками-монополистами, а также введение санкций в случае замедления доставки. Получение лицензий на авиаперевозки также следует увязать с закреплением минимальных лимитов перевозки почты.

4. Нормативно следовало бы закрепить максимальные контрольные сроки доставки корреспонденции почтовыми операторами связи (ФГУП «Почта России», «EMS Russian Post» и др.), а также предусмотреть отсутствующую сегодня материальную ответственность провайдеров за сохранность вложений. При этом солидарная материальная ответственность должна распространяться на всех участвующих в процессе доставки субподрядчиков (провайдеров транспортных услуг).

Таким образом, можно сделать вывод о том, что распределительная логистика находится в России на той стадии развития, когда определяющим фактором ее конкурентоспособности становится институциональная поддержка государства. Признание распределительной инфраструктуры в качестве стратегического ресурса модернизации российской экономики обусловлено сегодня самой логикой глобальной конкуренции. Принимаемые в этой области управленческие решения определят, станет Российская Федерация центром экономической интеграции постсоветского пространства или и дальше будет обречена на догоняющий тип развития.